\documentclass[12pt]{article}
\usepackage{times}
\usepackage{geometry}
\usepackage[utf8]{inputenc}
\usepackage{enumitem,amssymb}
\newlist{thematic}{itemize}{8}
\setlist[thematic]{label=$\square$}
\usepackage{pifont}

\usepackage{natbib,graphicx,amsmath}
\def\inv{$^{-1}$}
\newcommand{\la}{\lesssim}
\newcommand{\ga}{\gtrsim}

\usepackage{sectsty}
\sectionfont{\large}
\subsectionfont{\normalsize}

\newcommand{\presecsquish}{\vspace{-10pt}}  
\newcommand{\postsecsquish}{\vspace{-8pt}}    
\newcommand{\presubsecsquish}{\vspace{-8pt}}  
\newcommand{\postsubsecsquish}{\vspace{-4pt}}  

\newcommand{\apj}{ApJ}
\newcommand{\apjl}{ApJ}
\newcommand{\apjs}{ApJS}
\newcommand{\aap}{A \& A}

\newcommand{\mnras}{MNRAS}

\begin{document}
\thispagestyle{empty}
\begin{flushleft}
\huge
Astro2020 Science White Paper \linebreak

Detecting Offset Active Galactic Nuclei \linebreak
\normalsize

\noindent \textbf{Thematic Areas:} \hspace*{60pt} $\square$ Planetary Systems \hspace*{10pt} $\square$ Star and Planet Formation \hspace*{20pt}\linebreak
$\text{\rlap{$\checkmark$}}\square$ Formation and Evolution of Compact Objects \hspace*{31pt} 
$\text{\rlap{$\checkmark$}}\square$ Cosmology and Fundamental Physics \linebreak
  $\square$  Stars and Stellar Evolution \hspace*{1pt} $\square$ Resolved Stellar Populations and their Environments \hspace*{40pt} \linebreak
  $\text{\rlap{$\checkmark$}}\square$    Galaxy Evolution   \hspace*{45pt} 
     $\text{\rlap{$\checkmark$}}\square$        Multi-Messenger Astronomy and Astrophysics \hspace*{65pt} \linebreak
  
\textbf{Principal Author:}

Name: Laura Blecha	
 \linebreak						
Institution:  University of Florida, Gainesville, FL, USA
 \linebreak
Email: lblecha@ufl.edu	
 \linebreak
Phone:  352-392-4948
 \linebreak
 
\textbf{Co-authors:} Brisken, Walter,$^1$
Burke-Spolaor, Sarah,$^{2,3,4}$
Civano, Francesca,$^{5}$
Comerford, Julia,$^6$
Darling, Jeremy,$^6$ Lazio, T.~Joseph~W.,$^7$ and Maccarone, Thomas J. $^7$
\linebreak

{$^1$Long Baseline Observatory, Socorro, NM, USA; {wbrisken@nrao.edu}}\\
{$^2$Department of Physics and Astronomy, West Virginia University, P.O. Box 6315, Morgantown, WV 26506, USA}\\
{$^3$Center for Gravitational Waves and Cosmology, West Virginia University, Chestnut Ridge Research Building, Morgantown, WV 26505, USA}\\
{$^4$ CIFAR Azrieli Global Scholar; {sarah.spolaor@mail.wvu.edu}}\\
{$^5$ Harvard Smithsonian Center for Astrophysics, Cambridge, MA, USA; {fcivano@cfa.harvard.edu}}\\
{$^6$University of Colorado, Boulder, CO, USA; {julie.comerford@colorado.edu, jeremy.darling@colorado.edu}}\\
{$^7$Jet Propulsion Laboratory, California Institute of Technology, Pasadena, CA USA; {Joseph.Lazio@jpl.nasa.gov}}\\
$^8$Texas Tech University, Lubbock, TX, USA; {thomas.maccarone@ttu.edu}
  \linebreak

\textbf{Abstract:}
Gravitational wave (GW) and gravitational slingshot recoil kicks, which are natural products of SMBH evolution in merging galaxies, can produce active galactic ``nuclei'' that are offset from the centers of their host galaxies. Detections of offset AGN would provide key constraints on SMBH binary mass and spin evolution and on GW event rates. Although numerous offset AGN candidates have been identified, none have been definitively confirmed. Multi-wavelength observations with next-generation telescopes, including systematic large-area surveys,
will provide unprecedented opportunities to identify and confirm candidate offset AGN from sub-parsec to kiloparsec scales. We highlight ways in which these observations will open a new avenue for multi-messenger studies in the dawn of low-frequency ($\sim$ nHz - mHz) GW astronomy.
\end{flushleft}

\pagebreak
\setcounter{page}{1}
\presecsquish
\section{Offset AGN: Signposts for Supermassive Black Hole Mergers}
\postsecsquish

Supermassive black holes (SMBHs) are crucial components in the evolution of galaxies, and they are primary sources for low-frequency ($\sim$ nHz - mHz) gravitational wave (GW) observations with pulsar timing arrays (PTAs) and the LISA mission. We highlight the capabilities  of next-generation telescopes
to probe a poorly understood aspect of supermassive black hole evolution: how often are SMBHs offset from the nuclei of their host galaxies? Despite our nomenclature for active galactic nuclei, or AGN, SMBHs are not always centrally located. In ongoing galaxy mergers, for example, offset AGN occur when one of the SMBHs is active; many such objects have been found \citep[e.g.,][]{barrow16}. Here we focus on other types of offset SMBHs: gravitational-wave (GW) recoil and gravitational slingshot kicks, which can displace SMBHs or even eject them from galaxies entirely. Such systems are more than a mere curiosity. In addition to their effects on SMBH-galaxy evolution \citep[e.g.,][]{volont07,blecha11}, confirmed recoiling SMBHs can constrain the mass and spin evolution of binary SMBHs. This will inform the rate and characteristics of GW sources detectable with PTAs and LISA.

\presubsecsquish
\subsection{Recoiling Supermassive Black Holes}
\begin{figure*}
\centering
\includegraphics[width=0.2\textwidth]{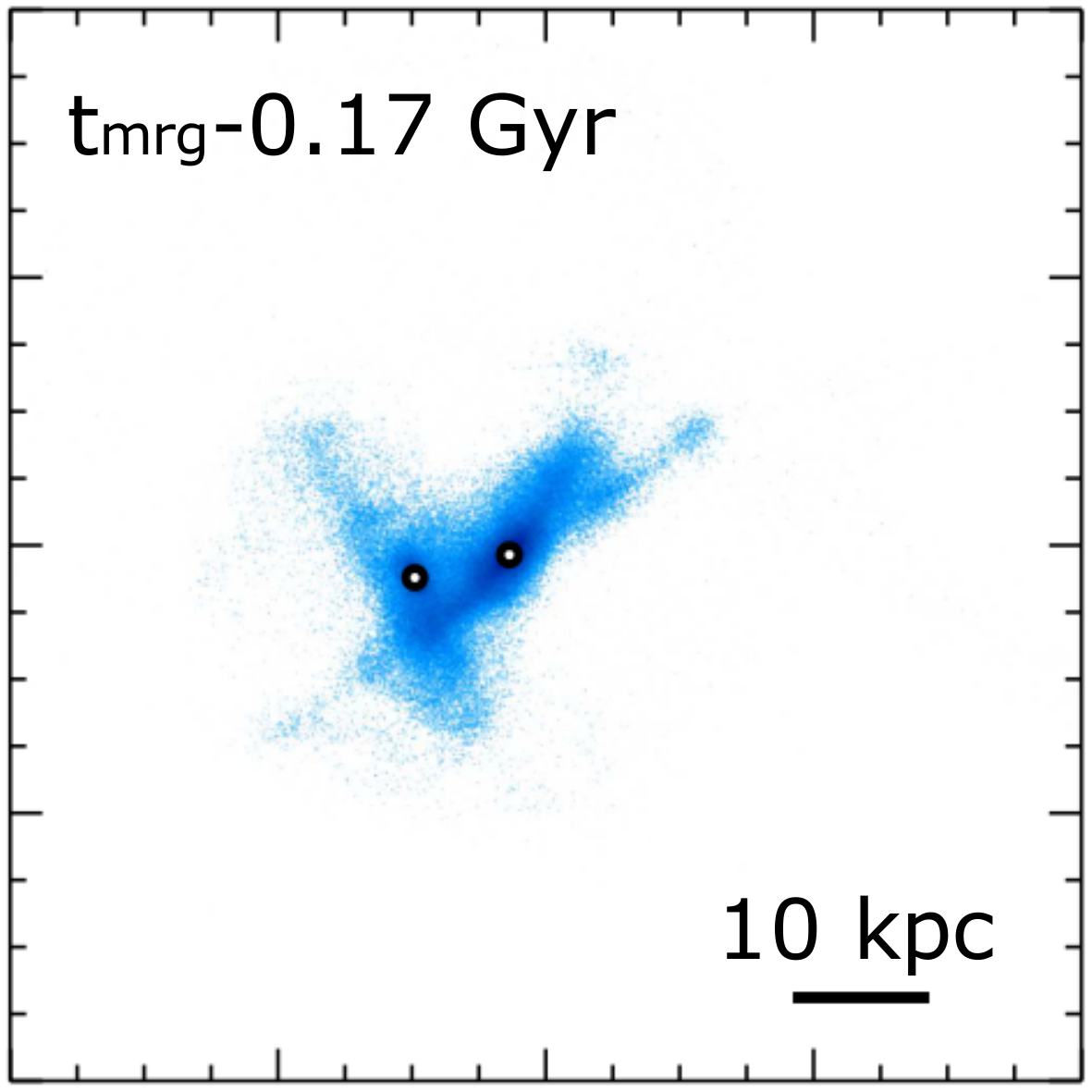}
\includegraphics[width=0.2\textwidth]{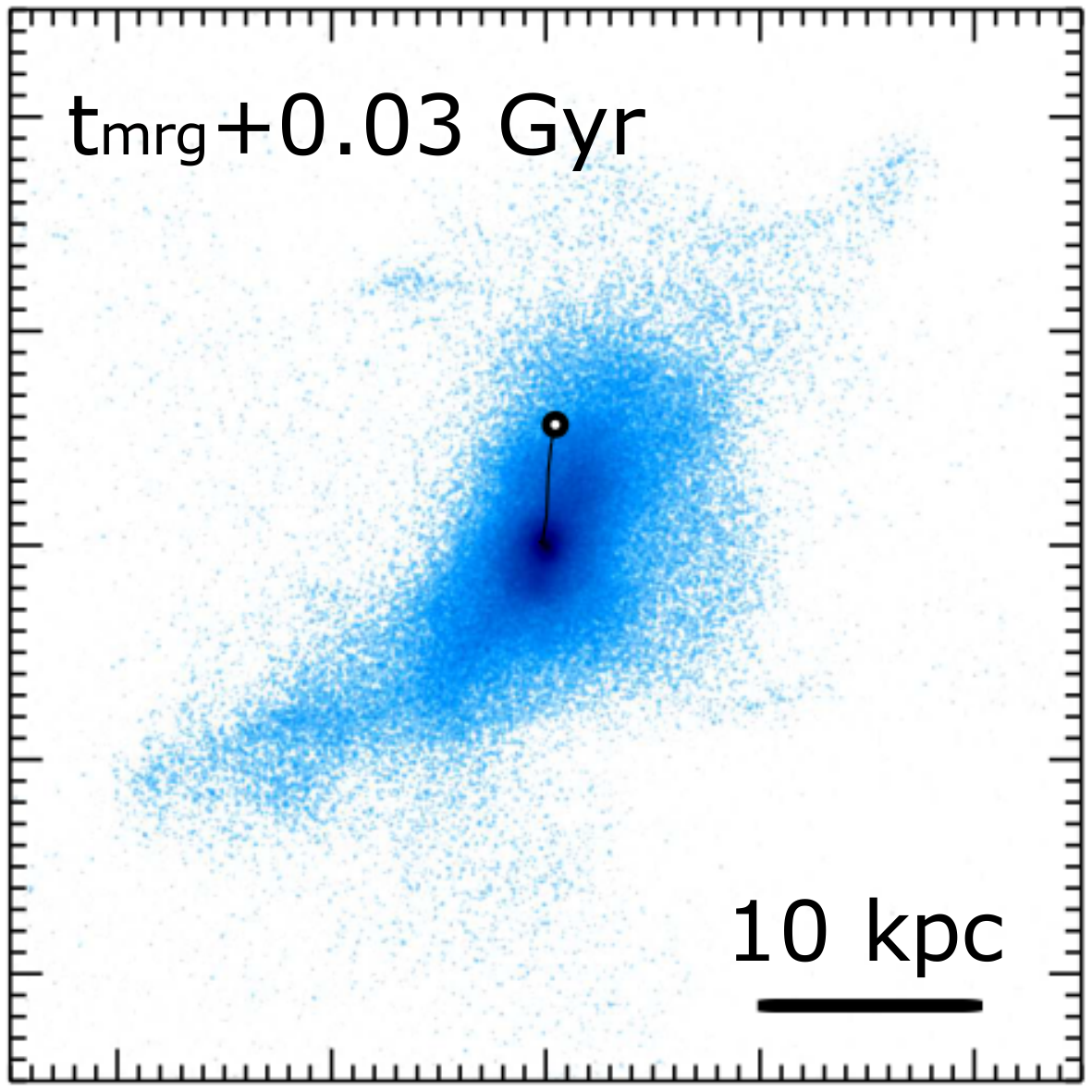}
\includegraphics[width=0.2\textwidth]{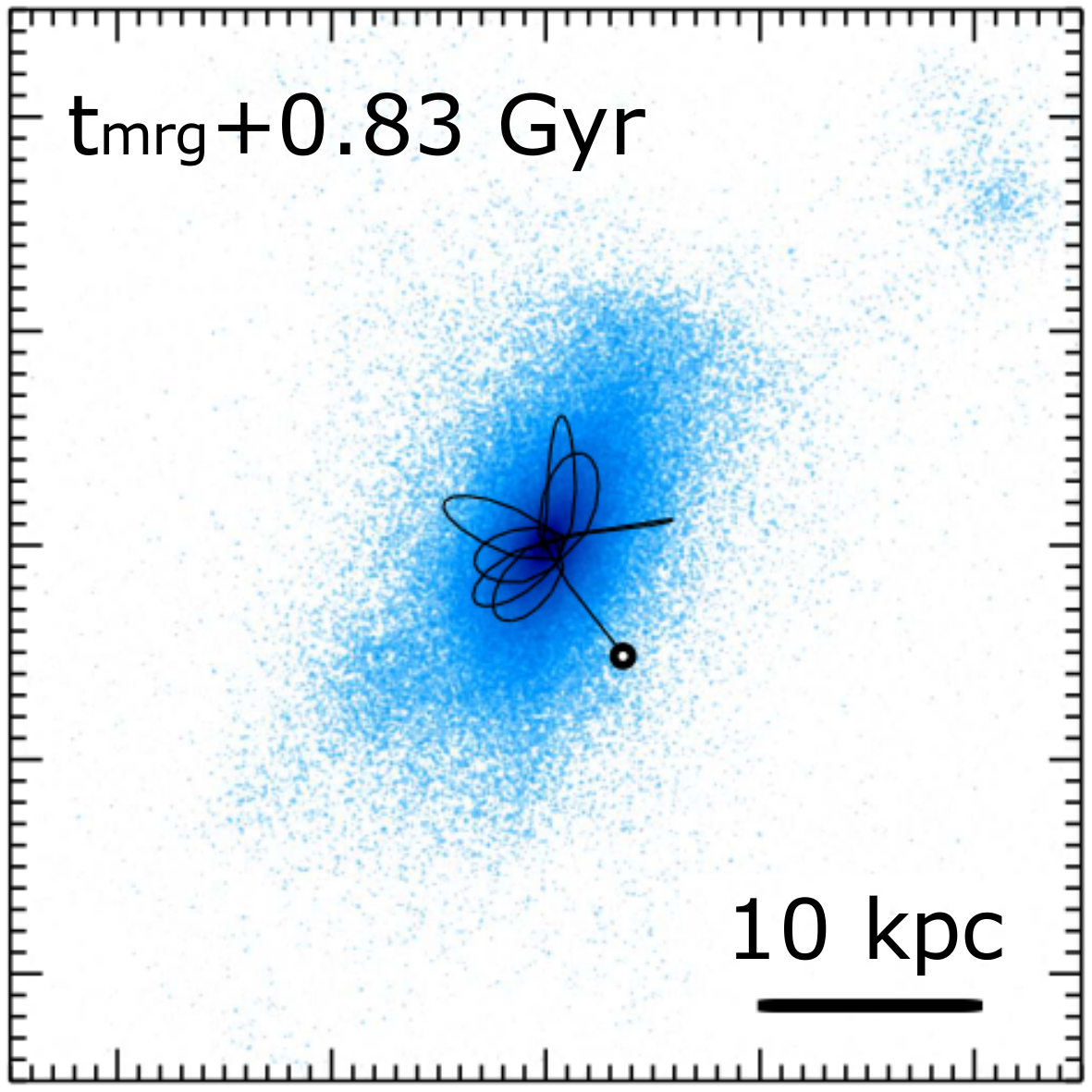}
\includegraphics[width=0.2\textwidth]{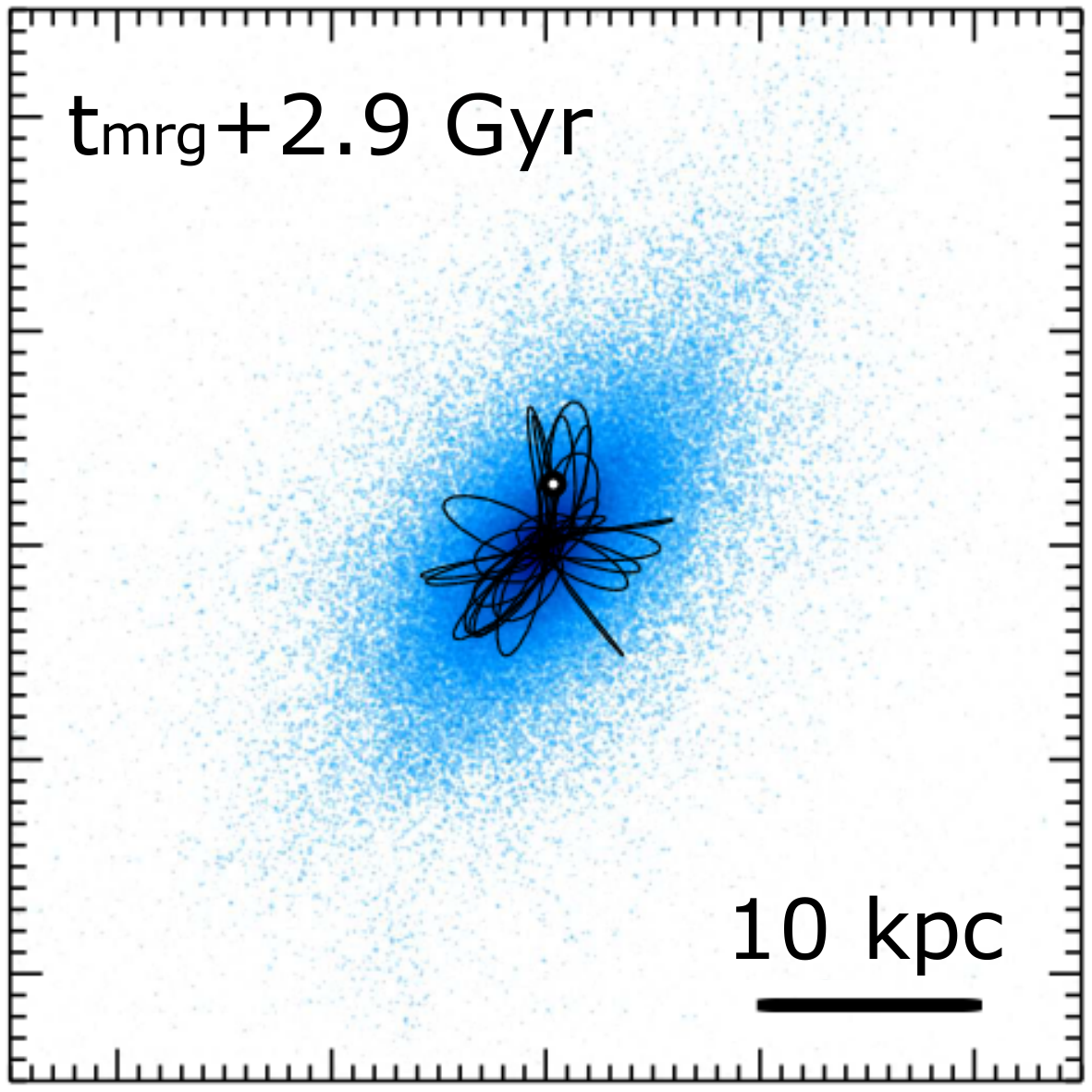}
\caption{Illustrative example of a galaxy merger and GW recoil from a hydrodynamics simulation \citep[cf.][]{blecha11}. The stellar density is shown in blue, and black dots denote the SMBH position. The left panel shows the two galaxy nuclei shortly before coalescence; subsequent panels show the recoiling SMBH trajectory after being kicked at 80\% of the central escape speed. \label{fig:recoil_traj}}
\end{figure*}

\postsubsecsquish
Asymmetric GW emission during a SMBH merger can impart a kick of up to 5000~km~s\inv\ to the SMBH merger remnant \citep{campan07b, lousto10}. 
Extreme recoil kicks should be exceedingly rare, but kicks of even a few hundred km s\inv\ can 
produce detectable offsets (e.g., \citealt{guamer08}, \citealt{bleloe08}; Figure \ref{fig:recoil_traj}). 
Because GW recoil velocities decrease sharply if SMBH spins are aligned before merger,
detections of offset, rapidly-recoiling SMBHs would provide strong evidence for misaligned, spinning progenitor SMBHs.

Alternately, if binary SMBH inspiral timescales are longer than the time between galaxy mergers, a subsequent merger may introduce a third SMBH into a galaxy. 
A three-body encounter will often eject the lightest SMBH, producing a gravitational slingshot kick and possibly driving the remnant binary SMBH to rapid merger \citep[e.g.,][]{hofloe07,bonett16,bonett18}. A key distinction between GW and slingshot recoil is that the latter leaves more than one SMBH remaining in the system after the kick.

\presubsecsquish
\subsection{Observable Signatures of Offset AGN}
\postsubsecsquish
In either case, if a recoiling SMBH is accreting at the time of the kick, it will carry along its accretion disk, broad emission line region, and radio-emitting core (everything within $\sim 10^{4}$--$10^{5}$ gravitational radii will typically remain bound to the SMBH). 
The recoiling SMBH could then be observed as an ``offset AGN'' with spatial and/or velocity offsets for up to tens of Myr \citep[e.g.,][]{madqua04,loeb07,blecha11}. 
However, unambiguous confirmation of recoiling AGN has proven challenging due to a number of factors, including the large samples needed for a discovery,  nuclear obscuration, possible confusion with an inspiraling pre-merger AGN, and the limited resolution and sensitivity of current instruments.

Very long baseline interferometry (VLBI) capable of sub-pc resolution in the radio (with, e.g., the ngVLA) will present an unparalleled opportunity to identify small spatial offsets. Here, the sensitivity to GW-recoiling AGN will be limited primarily by the relative astrometric accuracy of optical or infrared (IR) imaging of the host galaxy centroid. At least in older stellar populations, near-IR (NIR) astrometric centroiding  can be quite accurate \citep[within $\sim 100$ mas; cf.][]{condon17}. 
Thirty-meter  telescopes (e.g., ELT, GMT, TMT \& LSST) will enable even more accurate astrometry of optical galactic nuclei. Moreover, a slingshot recoil (where at least one SMBH remains behind as a secondary
 radio source), could be resolved down to 
 $\la$ 1\% of the beam FWHM for moderately bright sources. Next-generation very long baseline interferometry also offers the entirely new possibility of measuring proper motions of rapidly-recoiling AGN in nearby systems. 

\presecsquish
\section{Current Searches for Offset AGN and Limitations}
\postsecsquish

\begin{figure}
\centering
\includegraphics[width=0.27\textwidth]{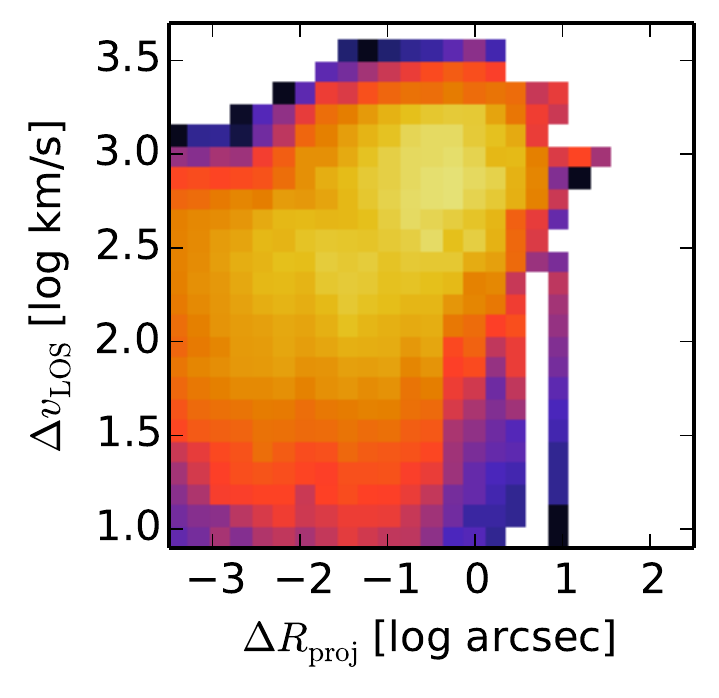}
\includegraphics[width=0.27\textwidth]{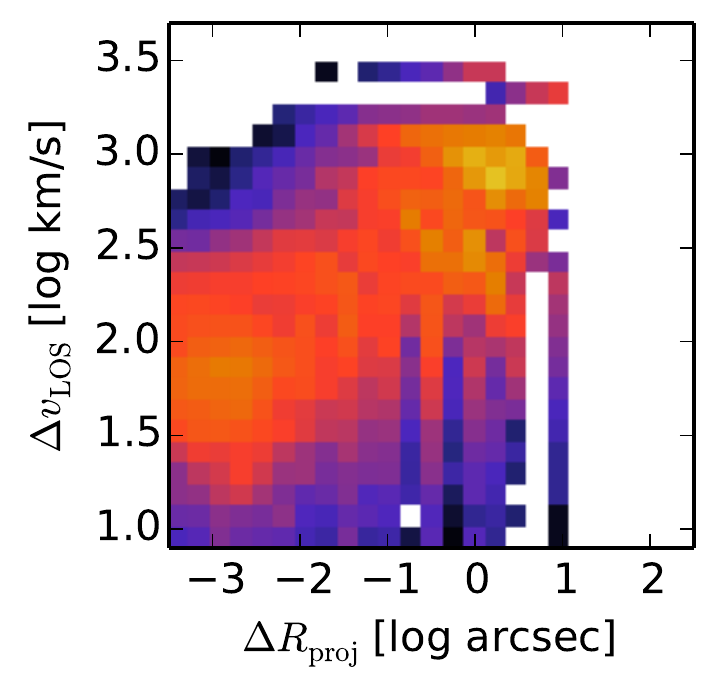}
\includegraphics[width=0.375\textwidth,trim = 6 0 0 6]{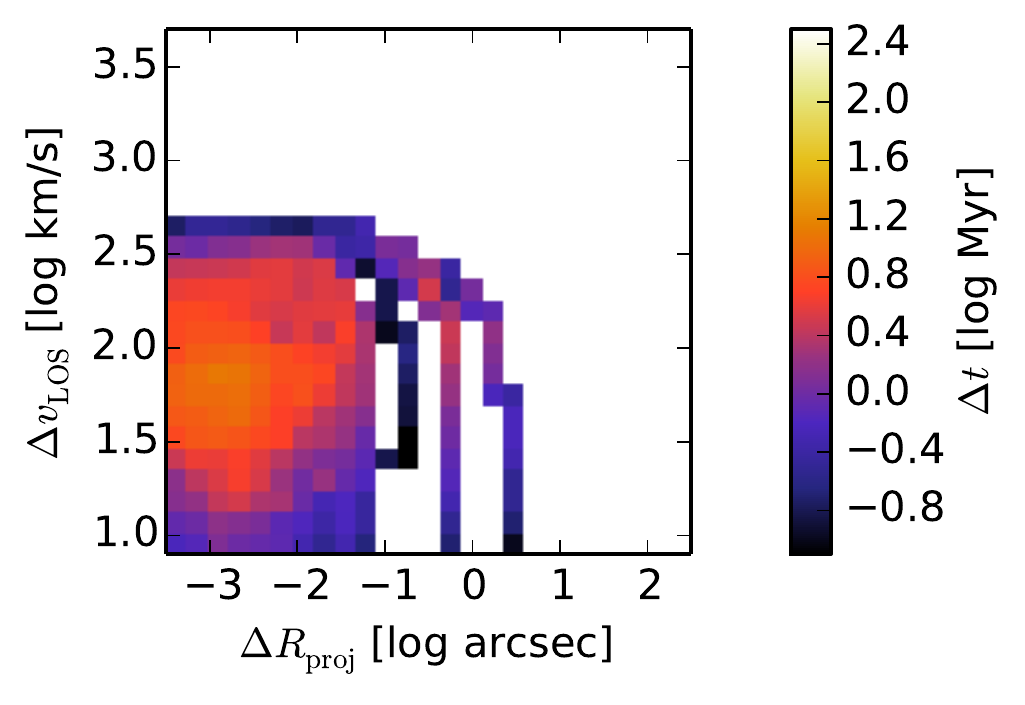}
\caption{Offset GW-recoiling AGN distributions, from the Illustris cosmological simulations and the models of \citet{blecha16}. 
{\em Left:} Pre-merger spins are randomly oriented.
{\em Middle:}  SMBH spins are aligned only in gas-rich mergers. 
{\em Right:} Pre-merger SMBH spins are aligned to within $5^{\circ}$. 
Only AGN that would be detected in {\em HST}-COSMOS are included. Upcoming large surveys in the optical, NIR, radio, and X-ray could detect hundreds of offset AGN, if spins are not always aligned.
\label{fig:offsets}}
\end{figure}

\subsection{Spectroscopic and kpc-scale Spatial Offsets}
\begin{figure}[tb]
\centering
\includegraphics[width=0.445\textwidth]{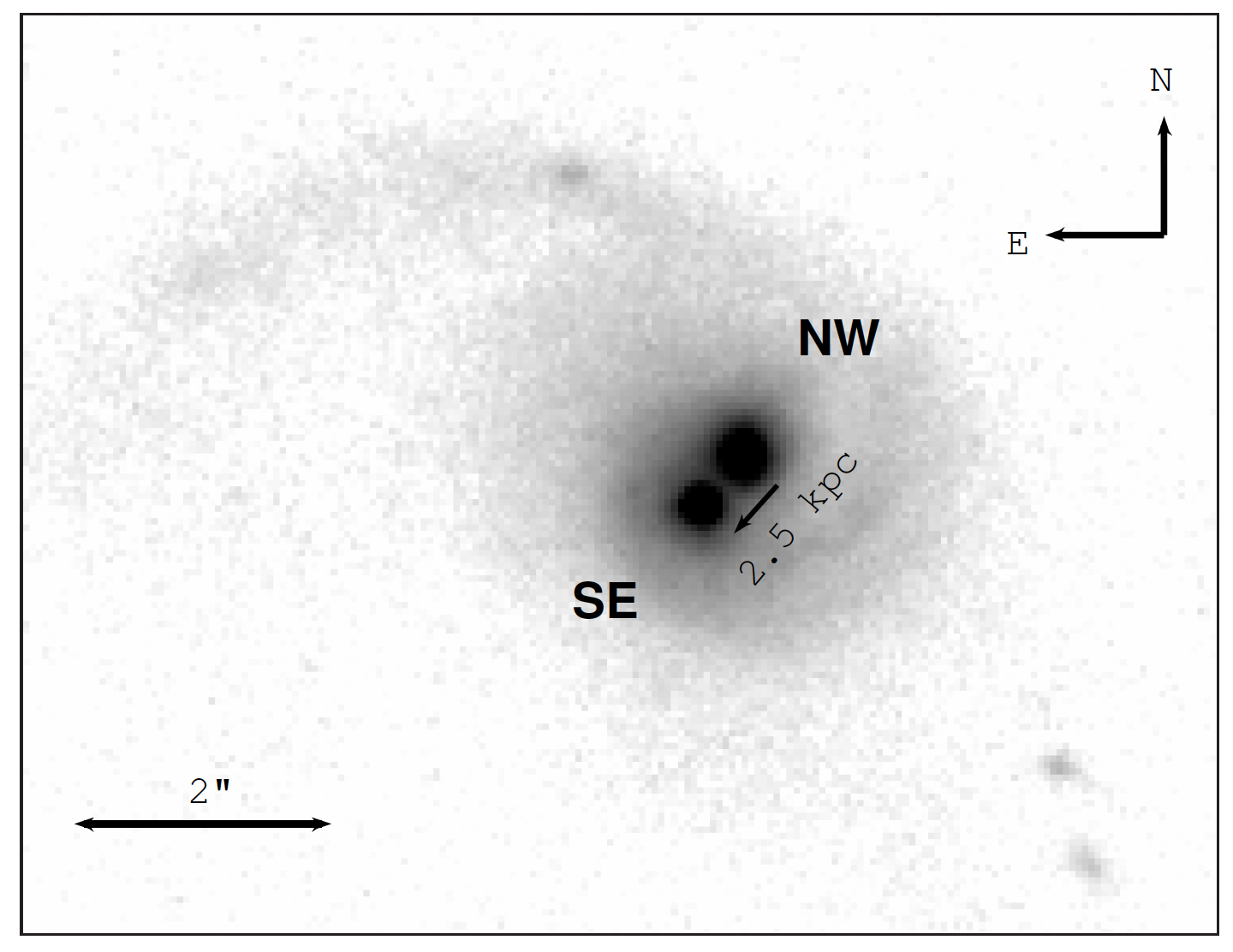}
\includegraphics[width=0.47\textwidth,trim={0cm 0cm 0cm 5cm},clip]{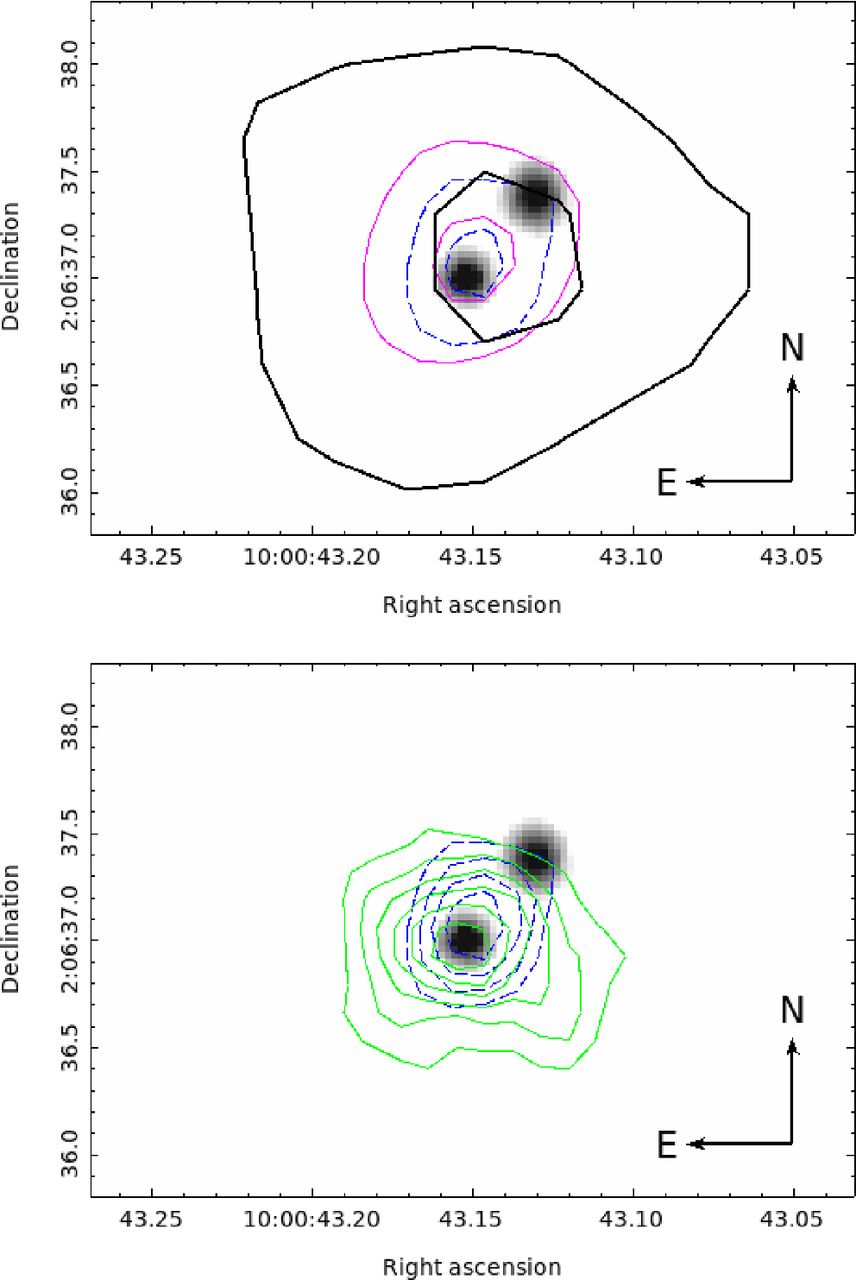}
\includegraphics[width=0.7\textwidth]{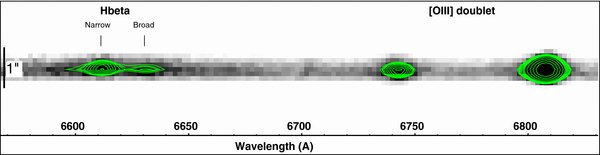}
\caption{The recoiling AGN candidate CID-42. {\em Top left:} The F814W {\em HST} image of this source shows a recent merger.
(From \citealt{civano10}.) {\em Top right:} Combined image of the CID-42 nucleus including HST optical (gray), Chandra X-ray (green), and VLA 3 GHz radio (blue). The X-ray and radio emission are consistent with a point source in the SE nucleus (the putative recoiling AGN). From \citealt{civano12} \& \citealt{novak15}. {\em Bottom:} Magellan/IMACS spectrum of CID-42, with a 1300 km s\inv\ offset between the broad and narrow H$\beta$ \citep{civano10}. \label{fig:candidates}}
\end{figure}

\postsubsecsquish
Numerous candidate recoiling AGN have been found. Some have been identified spectroscopically via Doppler-shifted optical broad lines \citep[e.g.,][]{ komoss08, robins10, eracle12}, and 
others have been identified via kpc-scale spatial offsets \citep{jonker10, koss14, markak15,kalfou17}. Alternate possibilities such as AGN outflows, binary SMBHs, or dual AGN are often difficult to exclude, however. 
The most promising candidates are those with both spatial and velocity offset signatures \citep{civano10, civano12, blecha13a, chiabe17}; CID-42 is one such example (Figure \ref{fig:candidates}). 
Still, current data cannot exclude an inspiraling, kpc-scale SMBH pair in which one SMBH is quiescent or intrinsically faint. To date, none of the recoil candidates have been confirmed. 
Future high-sensitivity, high resolution observations with the ngVLA, JWST, WFIRST, Euclid, thirty-meter optical telescopes, or X-ray missions such as Lynx or AXIS will provide much stronger constraints on AGN offsets and on the possible presence of a secondary, faint AGN in the host nucleus.

\presubsecsquish
\subsection{Parsec-scale Spatial Offsets}
\postsubsecsquish
If a recoiling SMBH is not ejected entirely from its host galaxy, it will eventually return to the nucleus,
where it may undergo long-lived, small-amplitude oscillations \citep[e.g.,][]{guamer08, bleloe08}. 
Several AGN with parsec-scale optical photometric offsets have been identified in nearby core ellipticals \citep{batche10,lena14}. However, AGN jet activity may be responsible for some apparent offsets \citep{loppri18}.
In order to determine the true prevalence of small-scale AGN offsets, higher-sensitivity radio observations are needed to probe low-luminosity and radio-quiet AGN in galaxies with diverse morphologies. High-resolution optical or NIR imaging of the host stellar light is  also needed to measure the position of the host centroid and to identify signatures of a recent merger (tidal tails, shells, etc.); this includes ground-based adaptive optics imaging with current instruments and future thirty-meter telescopes, as well as with JWST. For nearby galaxies, these optical and NIR observations alone will be able to resolve offsets down to $\sim$ parsec scales.

\presecsquish
\section{Required Observing Capabilities}
\postsecsquish
Identifying a population of offset AGN will require large surveys of thousands of galaxies that leverage the complementary strengths of radio, IR, optical, and X-ray telescopes. Key survey capabilities include high spatial resolution, high sensitivity, and large area coverage. These must be combined with multi-wavelength, multi-epoch broadband and spectroscopic follow-up to confirm the true nature of candidate recoiling SMBHs.

\presubsecsquish
\subsection{Surveys for Spatially- and Spectroscopically-offset AGN}
\postsubsecsquish

\citet{blecha16} predict that even if SMBH spins are largely aligned in gas-rich systems (where circumbinary disks can torque spins into alignment), hundreds of spatially offset AGN may be detected with large-area optical and NIR surveys using, e.g., LSST, WFIRST, and Euclid \citep[see also][]{volmad08}. Notably, LSST will utilize the time domain to distinguish stochastically-variable offset AGN from transient offset sources \citep[i.e., SNe; cf.][]{kumar15}. 

Surveys in the IR (with JWST, WFIRST, and Euclid) and in the radio (with, e.g., the ngVLA) will be uniquely sensitive to offset AGN embedded in obscured nuclei, which are common in merger remnants. Wide-field X-ray surveys with, e.g., Lynx or AXIS could detect moderately obscured offset AGN.
A VLBI survey, in particular, could detect AGN with $<$ pc - kpc-scale offsets; such a survey would also be well-aligned with the goals of a search for $\sim$ pc-scale and sub-pc-scale {\em binary} SMBHs \citep[e.g.,][]{burkes18}. 

If spins {\em are} always efficiently aligned to within a few degrees, GW recoil velocities are $\la 300$ km s\inv, and 
spatial offsets $\ga 0.05"$ are rare (Figure \ref{fig:offsets}). In this case, VLBI is about the {\em only} means of detecting GW recoils.
 The spatial offsets of GW-recoiling AGN would therefore constrain the pre-merger spins of SMBH binaries, which is otherwise very difficult prior to GW detections.

Nearly 100 quasars with large ($>1000$ km s$^{-1}$) broad line offsets have been identified in SDSS \citep[e.g.,][]{eracle12}. Many more such objects will be identifiable in newer optical spectroscopic surveys (e.g., SDSS eBOSS, SDSS-V, and DESI). Some of these may arise from the bulk motion of a recoiling SMBH, although the large majority are likely produced by other phenomena such as gas outflows, unusual double-peaked emitters, or even binary SMBHs \citep[e.g.,][]{decarl14,runnoe17}. Broad line velocity offsets observed in conjunction with spatial offsets and disturbed morphology indicative of a recent merger will provide compelling ancillary evidence for a recoiling AGN  (cf.\ CID-42, Figure~\ref{fig:candidates}), and the ``lowest-hanging fruit" would be AGN spatially offset by several kpc from the galactic nucleus \citep[e.g.,][]{blecha16}.

\presubsecsquish
\subsection{Follow-up of Recoiling AGN Candidates}
\postsubsecsquish

Multi-wavelength detection of an offset AGN, ideally from radio to X-ray wavelengths, is needed to confirm the AGN nature of the source. High-resolution X-ray observations with, e.g., Lynx or AXIS will be important for identifying radio-quiet AGN with faint radio cores. High dynamic range, high sensitivity observations will often be needed to distinguish between a kpc-scale dual AGN (in which one AGN is faint) and a single, offset recoiling AGN (cf.~Figure~\ref{fig:candidates}). High resolution optical/NIR imaging is also key for centroiding the host galaxy and confirming the offset. 

Variability studies (via LSST, or with dedicated radio, optical, or X-ray monitoring campaigns) will be important for confirmation of recoiling AGN candidates identified in single-epoch data. Radio variability can also be used to rule out apparent spatial offsets produced by transient jet phenomena \citep[cf.][]{loppri18}. Here, high resolution coupled with wide frequency coverage is ideal for distinguishing an offset AGN core from other features, such as a knot in a jet. Specifically, spectral index measurements can distinguish between a flat-spectrum core (indicating ongoing AGN activity) and a steep-spectrum jet (with an older electron population).

Resolved spectroscopy is crucial for pinpointing the location of AGN emission lines relative to the galactic nucleus. In addition, integral field spectroscopy will reveal dynamics of galactic nuclei, in order to discriminate between inspiraling kpc-scale dual AGN and a post-merger, recoiling offset AGN. For spectroscopically-offset AGN, VLBI will be able to resolve spatial offsets on pc to kpc scales corresponding to either GW-recoiling AGN (with one radio core), or slingshot recoil or binary SMBHs (with two radio cores). The latter could be distinguished with monitoring to identify possible periodic signatures indicating binary motion.

\presubsecsquish
\subsection{Proper Motion Measurements}
\postsubsecsquish

VLBI with a $\ga$ 1000 km baseline, such as is proposed for the long-baseline ngVLA, would be uniquely capable of providing proper motion measurements of recoiling AGN. Relative astrometric precision of $< 1$\% of the beam FWHM would enable proper motions of $\sim 1 \mu$as yr\inv\ to be detected out to $\sim 200$ Mpc over a 5--10~yr time baseline, for transverse velocities $\ga 1000$~km~s\inv. Based on the \citet{blecha16} models, we predict up to $\sim 10$ detectable recoiling AGN with such a transverse velocity within this volume.
In practice, achieving this for single objects (i.e., GW-recoiling AGN) will require constraints on secular galactic and jet motion using nearby sources and repeated multi-band observations.

Slingshot recoils resulting from triple SMBH interactions present easier targets for proper motion studies, as the SMBH(s) remaining in the galaxy nucleus would allow precise relative astrometry. 
Although the prevalence of slingshot recoils is not known, theoretical models suggest that triple SMBHs are not uncommon \citep[][]{kulloe12,kelley17a}.

\presubsecsquish
\subsection{LISA Electromagnetic Counterparts}
\postsubsecsquish

Because LISA will rarely give adequate time to localize EM counterparts of GWs from merging SMBHs beforehand, a technique for identifying the EM counterparts is essential for identifying their host galaxies and doing precision cosmology. 
The SMBH merger product will be less massive that the sum of the
two progenitor SMBHs, such that the accretion disk will temporarily fail to reach the innermost stable circular orbit (ISCO) of the new system and the jet emission will falter, starting at high radio frequencies and gradually progressing to lower frequencies. 
The amplitude of this radio variability should be far larger than typically seen in AGN and thus could be identified with monitoring on $\sim$ weeks to years timescales. 
In addition, recent mergers of SMBHs with misaligned spins could produce multiple radio hot spots.

\presubsecsquish
\subsection{Other Multi-messenger Synergies with LISA and PTAs}
\postsubsecsquish

Any detections of recoiling (or binary) SMBHs in advance of LISA science observations can be used to constrain the LISA event rate. Moreover, even non-detections will constrain pre-merger spin evolution and thus the likelihood that LISA will observe precessing waveforms. PTAs may detect GWs from single SMBH binaries in the coming decade, possibly even before they detect the stochastic GW background \citep[e.g.,][]{kelley18}. 
Thus, the multi-wavelength EM searches for offset AGN described here, combined with GW searches for SMBHs binaries and mergers with PTAs and LISA, will transform our understanding of SMBH evolution in merging galaxies---a critical issue for low-frequency GW astronomy in the coming decades.

\pagebreak

\end{document}